# Econophysics of precious stones

A. Watanabe[1], N. Uchida[2] & N. Kikuchi[3]

[1]*3-55-1 Hatagaya, Shibuya, Tokyo 151-0072, Japan*

[2]*Department of Physics, Tohoku University, Sendai 980-8578, Japan*

[3]*Fachbereich Physik, Martin-Luther-Universität Halle-Wittenberg, D-06099 Halle, Germany*

**The importance of the power law has been well realized in econophysics over the last decade[1]. For instance, the distribution of the rate of stock price variation[2] and of personal assets[3] show the power law. While these results reveal the striking scale invariance of financial markets, the behaviour of price in real economy is less known in spite of its extreme importance. As an example of markets in real economy, here we take up the price of precious stones which increases with size while the amount of their production rapidly decreases with size. We show for the first time that the price of natural precious stones (quartz crystal ball, gemstones such as diamond, emerald, and sapphire) as a function of weight obeys the power law. This indicates that the price is determined by the same evaluation measure for different sizes. Our results demonstrate that not only the distribution of an economical observable but also the price itself obeys the power law. We anticipate our findings to be a starting point for the quantitative study of scale invariance in real economy. While the Black–Sholes model provided the framework for optimal pricing in financial markets [4], our method of analysis prvides a new framework that characterizes the market in real economy.**

We took the price data of precious stones from on-line jewelers, considering their public accessibility. First we analyse the selling price of diamonds published in a database, which takes the price lists from a number of jewelers [5]. We should note that the price of diamonds depend not only on the weight (measured in unit of Carat: 1 Ct = 0.2 gram) but also on the clarity, colour, and cut, according to which the stones are strictly categorized. To remove the dependence on the factors other than the weight, we use only the data of top quality (i.e. flawless and internally flawless, colour: D) diamonds with a round brilliant cut.

Fig.1 describes how the selling price $y$ scales with the weight $x$ in double logarithmic plot. We find that the price has power law dependence on weight: $y \sim x^{2.02 \pm 0.006}$. Here we stress that the exponent (slope of the curve), but not the prefactor of the power law, describes the trend of the graph. The exponent larger than unity means that the unit price (per weight) becomes highr for larger stones. If the exponent equals to 1, it means that the price is purely proportional to the weight, which is unlikely if we consider the rare production of large homogeneous crystals.

As examples of colour gemstones, we chose round blue sapphire and emerald, and plotted in Fig.2 their price as a function of weight. We find that they also obey the power law $y \sim x^a$ with the exponents $a_1 = 1.34 \pm 0.03$ (sapphire) and $a_2 = 1.38 \pm 0.07$ (emerald), which are very close to each other but clearly smaller than that of the diamond. Thus, the exponent depends on the type of precious stones, but it might take a close value for similar type of stones. It is suggested that the price of more precious stones like diamond has higher exponent. The scale invariance of the price (i.e. power law) for different sizes means that the same evaluation rule is used in the price determination. The origin of power law could be attributed to the balance between total cost of production and consumer's demand.

Fig.3 describes the price of quartz crystal ball for differnt weights. In this case the price range can be divided into three scaling regimes. For small weights, the exponent is $a_1 = 0.81 \pm 0.07$. It means that the unit price is higher for smaller stones, which might be due to the extra costs such as packaging or human resources. In the second regime with the price ranging from $10^4$ to $10^7$ JPY, the power law is very accurate with the exponent $a_2 = 1.53 \pm 0.007$. Also, the crossover from the first to second regime is quite sharp. At the end of the second regime the price suddenly becomes higher by a factor or about 3, resulting in discrete jump to the third regime, which is characterized by the exponent $a_3 = 1.51 \pm 0.025$. Note that the jump affects the prefactor of the power law, while the exponents for the two regimes are identical within the error bar. It is remarkable that a single exponent describes the trend of the graph for the price range of over 5 decades. The jump might be explained by the fact that the data for the third regime are for the crystal balls that are currently out of stock.

In Figure 4 we show the price of diamond at a retail company [8]. The data can be divided into 5 or 6 groups according to the weight and in each group the price exhibits a band structure, which reflects the difference in clarity and colour. Fitting to all the data points gives the exponent $a = 1.77 \pm 0.02$, while in each group, the price of top quality (internally flawless) samples obeys a power law with exponent close to 1, and jumps to the next group. The latter features suggest that the smooth power law behavior is strongly modified by the company. However, the overall power law tells that the market on a large-scale is determined by the balance between production and demand.

The method presented in this work may be usefully extended to quantitative analysis of other markets in real economy. The power law might be a sign of self-organization in a market with high fluidity and competition.

1. R. N. Mantegna and H. E. Stanley, "Introduction to Econophysics", Cambridge Univ. Press (2000).

2. R. N. Mantegna and H. E. Stanley, Nature **376**, 46-49 (1995).

3. H. Aoyama, W. Souma, Y. Nagahara, M. P. Okazaki, H. Takayasu and M.Takayasu, Fractals **8**, 293-300 (2000).

4. F. Black and M. Sholes, Journal of Political Economy, **81**, 637-654 (1973).

5. Diamond Review, http://www.diamond.info

6. IndyGem, Inc., http://www.indygem.com

7. Kanai Shouyaku, Co. Ltd., http://crystal-temple.com

8. ON-LINE-DIAMOND, Inc., http://www.on-line-diamond.com

**Acknowledgements**   This work is supported by ERYS of Tohoku university (N.U.) and DFG (N.K.).

**Figure 1**   The dependence of the diamond price (USD) on weight (Carat). 746 data points of top quality diamonds (flawless and internally flawless with colour type D with a round brilliant cut) are cited from an online diamond review [5]. The solid line is least square fit by the power law $y = bx^a$ to the data with slope $a = 2.02 \pm 0.006$ and prefactor $b = 4.12$.

**Figure 2**   The dependence of the colour gemstones' price (USD) on weight (Carat). The data of round blue sapphire (140 points) and emerald (88 points) are cited from IndyGem, Inc [6]. The solid lines are least square fits by the power law to the data with slopes $a_1 = 1.34 \pm 0.03$ (sapphire) and $a_2 = 1.38 \pm 0.07$ (emerald).

**Figure 3**   The dependence of the quartz crystal ball price (JPY) on weight (Carat). 134 data points of top quality crystal (natural crystal of 100% transparent, no cut, no inclusion) are cited from Kanai Shouyaku Co. Ltd[7]. The solid lines are least square fit by the power law $y = bx^a$ to the data with slopes $a_1 = 0.81 \pm 0.07$, $a_2 = 1.53 \pm 0.007$, and $a_3 = 1.51 \pm 0.025$ for the first, second and third regime.

**Figure 4**   The dependence of the diamond price (JPY) on weight (Carat). 705 data points of top quality diamonds (clarity: flawless to very very small inclusion, colour: D to H, with a round brilliant cut) is cited from ON-LINE-DIAMOND Inc[8]. The solid line is least square fit by the power law to the data with slope $a = 1.77 \pm 0.02$.

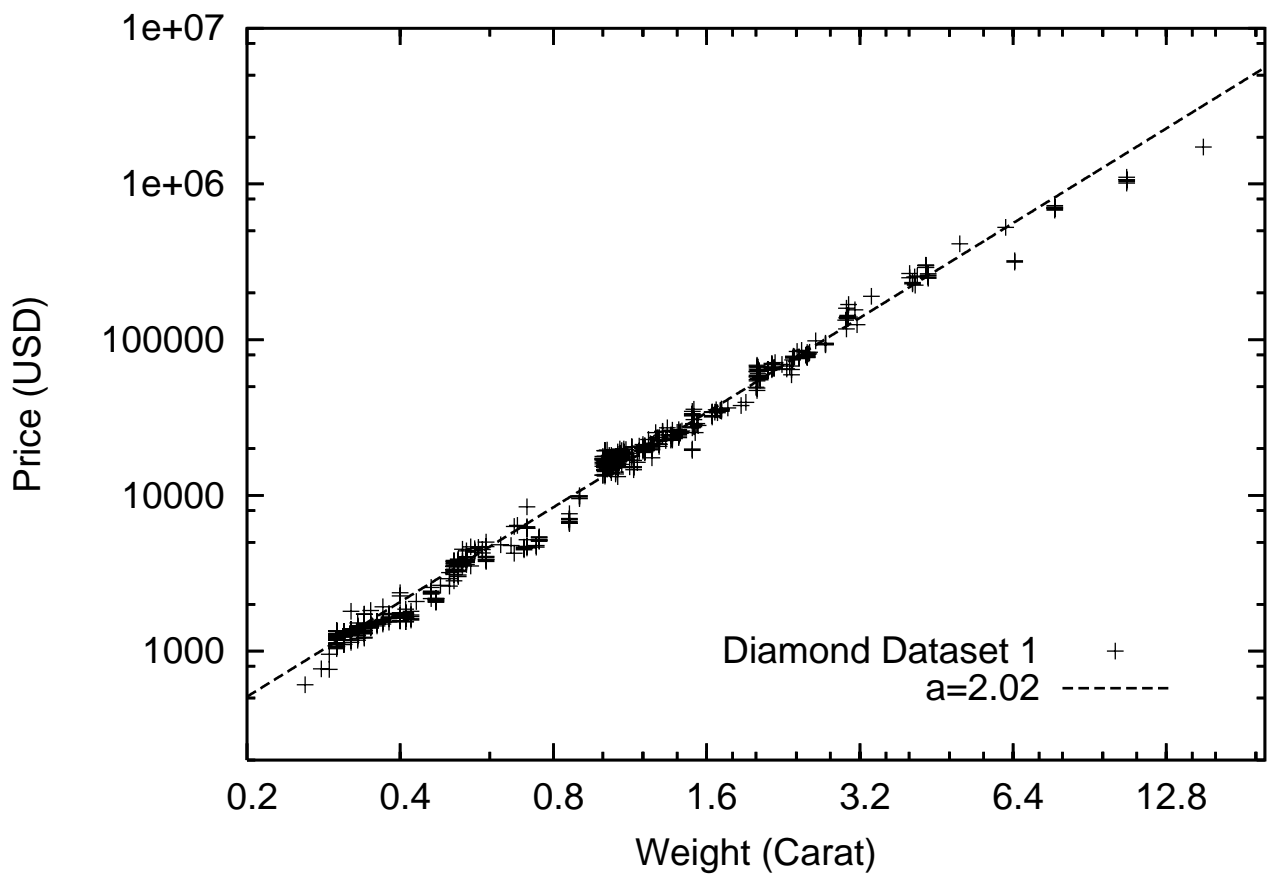

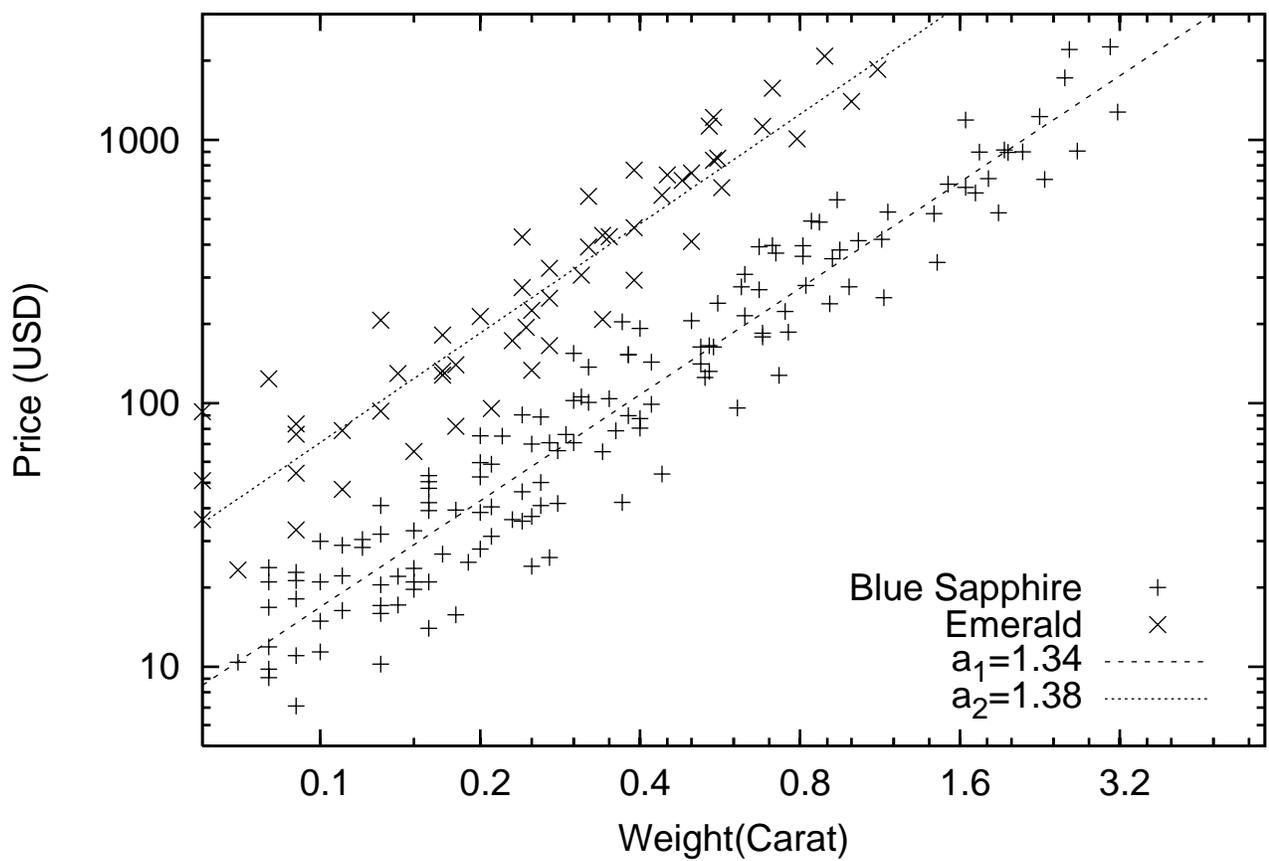

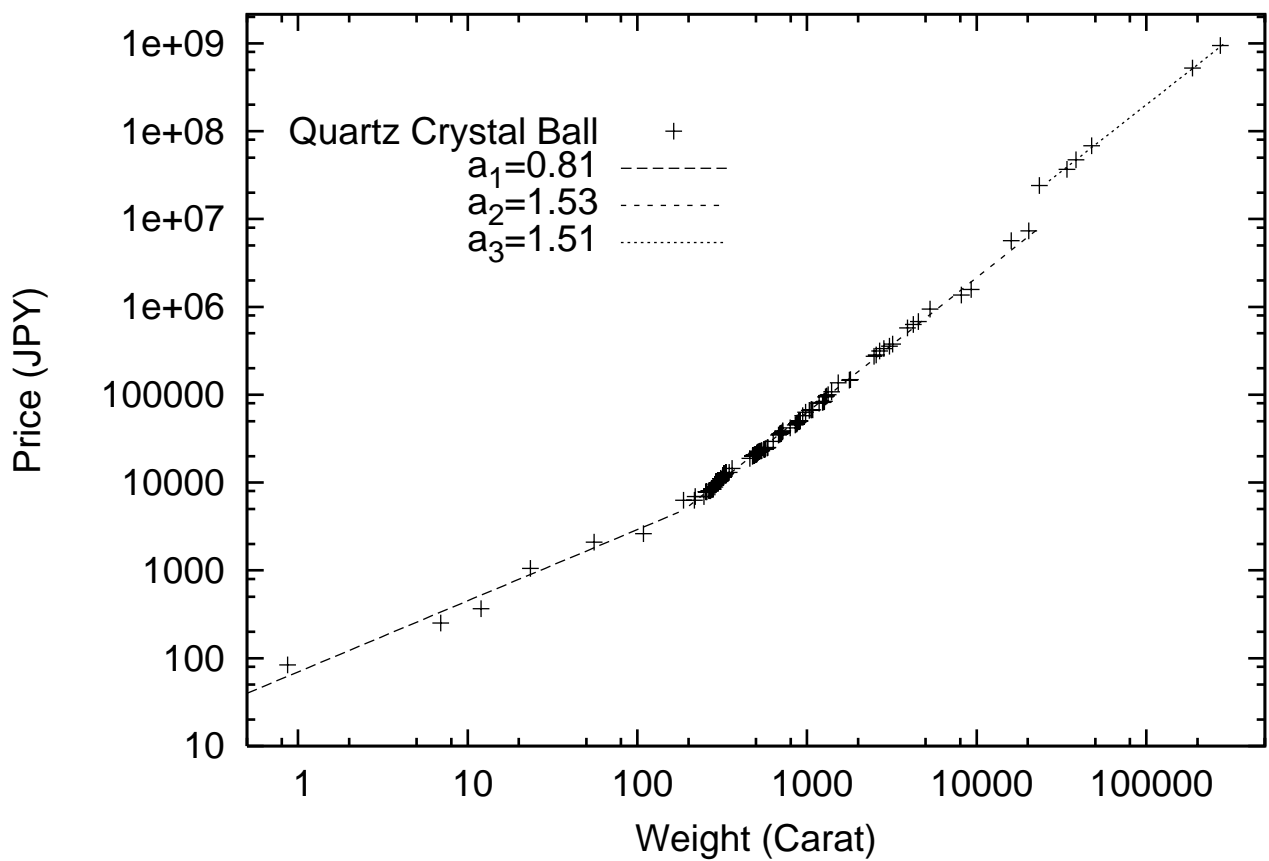

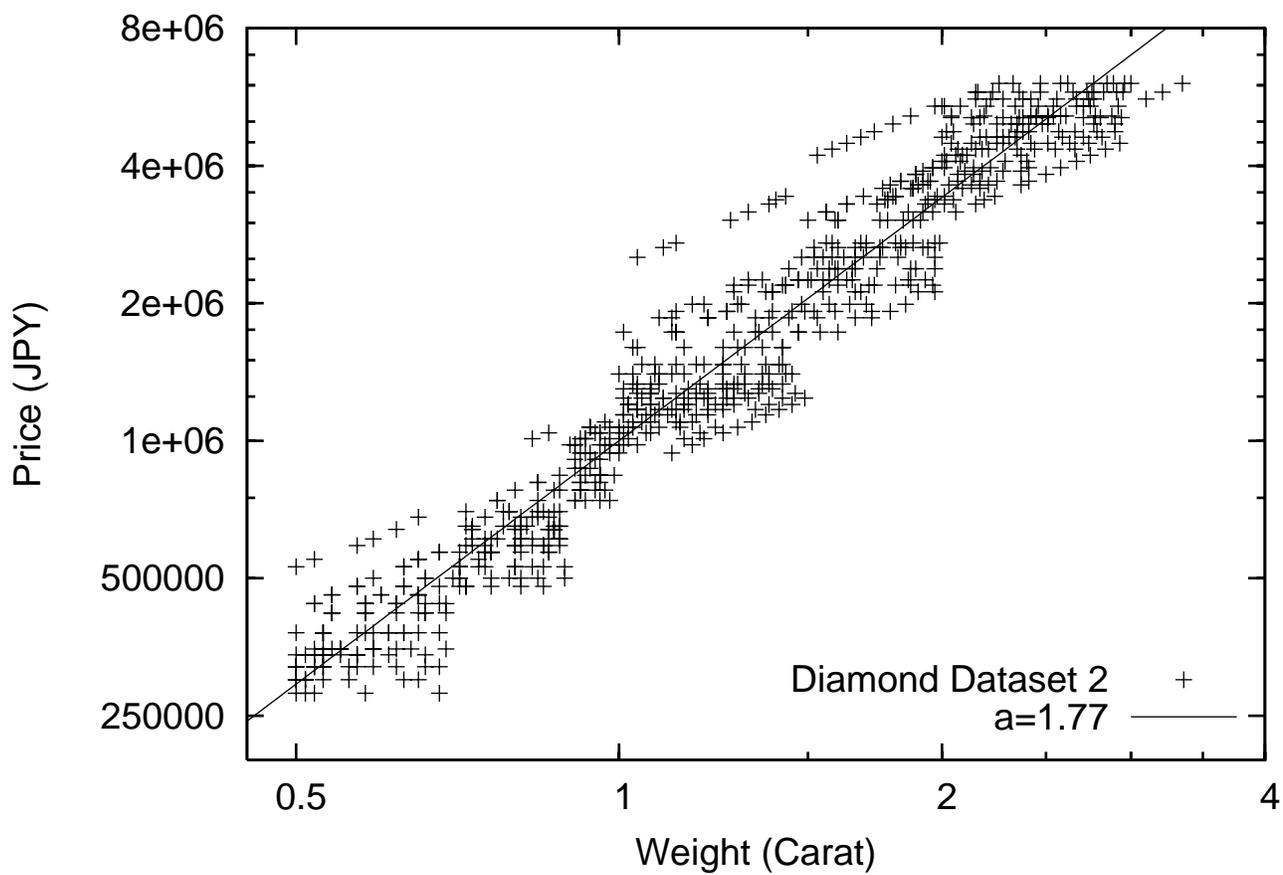